\documentstyle{article}
\begin{document}
\title{Supersymmetry and localization.    }
\author{Albert Schwarz\thanks{Research is
 partially supported by NSF grant No. DMS-9500704}
  ~and
Oleg Zaboronsky\thanks{On leave from the Insitute of  Theoretical and
Experimantal Physics, Moscow, Russia} \\
Department of Mathematics, University of California, Davis, CA 95616 \\
ASSCHWARZ@UCDMATH.UCDAVIS.EDU, \\
 ZABORON@UCDMATH.UCDAVIS.EDU}
\maketitle
\newtheorem{thm}{Theorem}
\newtheorem{lm}{Lemma}
\begin{abstract}
We study conditions under which an odd symmetry
 of the integrand leads to localization
of the corresponding integral over a
 (super)manifold.  We also show that in many cases
these conditions guarantee exactness of the stationary phase approximation
of such integrals.
\end{abstract}
\section{Introduction}

Localization formulae express certain integrals over (super)manifolds
as sums of contributions of some subsets of these manifolds.    Such
formulae were
studied in various contexts both in mathematics and physics.
Some important examples of localization formulae are based on
the theory of equivariant cohomology (see \cite{G} for a
review of the theory and its applications).    One famous particular
case is the Duistermaat-Heckman integration
formula \cite{DH} which became a powerful (though not completely
rigorous) tool in Quantum Field Theory (QFT).    In the context
of QFT, the Duistermaat-Heckman theorem gives sufficient conditions
for exactness of semiclassical approximation of field theoretical
models  (see
\cite{BT} ,\cite{W} and references
therein for a review of applications of Duistermaat-Heckman
and some other localization formulae to QFT).

The aim of the present paper is to derive very general
localization formulae in the framework of supergeometry.
Namely,  we consider an integral over a finite
 dimensional (super)manifold
$M$ where the integrand is invariant under the action of an
odd vector field $Q$.    We formulate sufficient conditions on $M$
and $Q$ under which the integral localizes onto the zero locus
of the number part of $Q$.
It is important to stress that without the
 conditions below, the localization
formula can be wrong.    (Physicists often used the localization formulae
without rigorous justification and without mentioning the conditions of
applicability of these formulae).

One of the possible ways to apply the theorems of the present
paper is based on the use of Batalin-Vilkovisky \cite{BV}
formalism where
the calculation of physical quantities reduces
 to the calculation of integrals
of functions invariant with respect to an odd vector field.
However the conditions
of our theorems are not always satisfied in this situation.

In the present paper we do not consider concrete examples of the
application of our results.  Let us mention only that they can be
used, for instance, to calculate
integrals of $OSp (n \mid m)$-invariant functions and obtain
the statement about
dimensional reduction in the
Parisi-Sourlas model  \cite{PS}
 and in similar situations \cite{P}.

To  conclude the introduction let us define a technically convenient
notion of compact vector field which will be used throughout the paper:
we say that a vector field $A$ on a (super)manifold $M$ is a
compact vector field if it generates an action of a one-parameter
subgroup of some compact group $G$ of transformations of $M$.
In other words we assume that there exists a homomorphism
$\varphi_{\ast}$ of the Lie algebra $\cal G$ of the compact Lie
group $G$ into the Lie algebra $Vect(M)$ of  vector fields on $M$
such that $ A \in Im \varphi_{\ast}$.

One can say also that the vector field $A$ on $M$ is compact
if the closure $G_{A}$ of the one-parameter group generated
by $A$ in $Diff (M)$ is compact.    Here $Diff (M)$ denotes
the group of diffeomorphisms of $M$ equipped with the compact
open toplogy (see \cite{Br}).    It is easy to see that
$G_{A}$ is a commutative connected compact Lie group.
Therefore it is isomorphic to a torus.
The space of compact vector fields on $M$ will be denoted by
${\cal K} (M)$.

Several definitions we use are explained in the Appendix.
All manifolds, maps, functions considered in this paper
are assumed to be smooth.

\section{The Duistermaat\-Heckman formula in the
language of supergeometry.}

First let us recall the conventional formulation
of the Duistermaat-Heckman integration formula (see
\cite{G} or \cite{BT} for a review).

Let $(W^{2n},  \Omega)$
be a compact $2n$-dimensional symplectic manifold
with a symplectic form $\Omega =
\Omega _{ij} (x) dx^i dx^j$.   Let $X \in {\cal K} (W)$
be a compact Hamiltonian vector field.
Let us denote
the corresponding Hamiltonian by $H$.
The Diustermaat-Heckman theorem states that
\begin{eqnarray}
\int_{W} \Omega ^n e^{iH} =\mbox{sum of contributions
of  the zero locus of the vector field $X$.   }
\end{eqnarray}
If in particular the zero locus $R$ of $X$   is a finite
subset of $W$ and all zeros of $X$ are non-degenerate
then
\begin{eqnarray}
\int_{W} \Omega ^{n} e^{iH} =i^{n}
\sum_{p \in R} e^{(i \frac {\pi }{4} sgn ~H (p))}
\frac{e^{iH (p)}}{\sqrt{\det Hess H (p)}},
\end{eqnarray}
where $Hess H (p)$ stands for the Hessian of $H$
at the point $p$ and
 $sgn~H (p)$ is the signature of $Hess H (p)$.

To show the relation of the Duistermaat-Heckman theorem
to supergeometry we notice, following \cite{W}, that the
left hand side of (1) can be rewritten as an integral over
a supermanifold.
Namely,  it is easy to check that
\begin{eqnarray}
\int_{W} \Omega ^{n} e^{iH} =i^{-n}
\int_{\Pi TW } \prod_{k=1}^{2n} dx^k d \xi ^k
e^{i(H (x) +\Omega _{ab} (x) \xi ^a \xi ^b )},
\end{eqnarray}
 where $\Pi TW$ denotes the total space of the tangent
bundle over $W$ with reversed parity of the fibers.   In other words if
$(x^{1},  \ldots ,  x^{2n})$ is a local coordinate system
in $W$, then a local coordinate system in $\Pi TW$
consists of even coordinates $(x^{1} ,  \ldots ,  x^{2n})$
and odd coordinates $(\xi ^{1} ,  \ldots ,   \xi ^{2n})$;
the coordinates  $(\xi ^{1} ,  \ldots ,   \xi ^{2n})$
transform as a vector by
the change of local coordinates $(x^1 ,  \ldots ,  x^{2n})$.

A function on $\Pi TW$ can be identified with
a differential form on $W$.   Using this remark
we can identify the exponential $S(x,  \xi )
\equiv H (x) + \Omega _{ij} (x) \xi ^i \xi ^j $
with the (inhomogeneous)
differential form $H +\Omega$.   Let us consider
the vector field $Q = \xi ^i \frac{\partial}{\partial x^i}
+X^{i} (x) \frac{\partial}{\partial \xi ^i}$.   As an operator acting
on forms on $M$,  $Q$ coincides  with the ``equivariant
differential'' $d+ i_{X}$,  where $i_{X}$
denotes the contraction with the vector field $X$.   It is easy to check
that the function $S(x,  \xi )$ on $\Pi TW$ satisfies $Q S =0$.
(The differential
form $H+\Omega$ obeys $(d+i_{X})(H +\Omega ) =0 $ because $X$ is
a Hamiltonian vector field with Hamiltonian $H$.   )

We conclude from this that the Duistermaat-Heckman theorem can be
reformulated in the following way:
\begin{eqnarray}
\int_{\Pi TW} dV e^S = \mbox{sum of contributions from
the zero locus of the vector field $Q$}.
\end{eqnarray}
Notice that $Q$ is an odd vector field on $\Pi TW$ such that
$Q^2 \in {\cal K} (\Pi TW)$ .   Let us explain this fact.
There exists a natural homomorphism
 of the algebra of vector fields on $W$
into the algebra of vector fields on $\Pi TW$.
(An infinitesimal diffeomorphism of $W$ induces
 an infinitesimal diffeomorphism
of $\Pi TW$.)
This homomorphism
transforms a vector field $X$ into
 the Lie derivative $L_{X}$,  considered as a
vector field on $\Pi TW$ (recall that the functions on $\Pi TW$ can be
considered as forms
on $W$;  therefore the Lie derivative can be regarded as a first order
differential operator acting
on functions on $\Pi TW$).   It is clear that when a vector field $A$
on $W$ generates an action of a subgroup of
a compact Lie group,  the corresponding vector field $L_{A}$ also generates
an action of a
subgroup of the compact Lie group on $\Pi TW$.
We identified $Q$ with $d + i_{X}$,  therefore
$Q^2$ is identified with $d i_{X} +i_{X} d =L_{X}$.
Thus we conclude that $Q^2 \in {\cal K} (\Pi TW)$.

In (4), $dV$ stands for $\prod_{i=1}^{2n} dx^i d \xi ^i$,
which is the canonical  volume element on
$\Pi TW$.   Note that
this volume element is $Q$-invariant;  i.e.,
the divergence of $Q$ with respect to
$dV$ vanishes: $div _{dV} Q =0$.(The notion of divergence is naturaly
generalized to the case of vector fields on an arbitrary supermanifold
 $M$ by the formula $\int_{M} dV (Q \cdot f) =
- \int_{M} dV (div_{dV} Q) f$.)

Eq.   (4) is equivalent to Eq.   (1) in virtue
of (3) and the one-to-one correspondence between sets of zeros of $Q$
and zeros of $X$.

It is natural to conjecture that Eq.   (4) remains correct if
$\Pi TW$ is replaced with an arbitrary supermanifold $M$,
$Q$ is an odd vector
field on $M$,  $dV$ is replaced with any $Q$-invariant
volume form on $M$,  and $S$ stands for any $Q$-invariant function.
We will see that this conjecture
is essentially correct if  $Q^2 \in {\cal K} (M)$; i.e.,  it is compact.

\section{Localization of integrals over supermanifolds.}

In this section we will formulate and prove
several statements giving sufficient conditions under which
an integral over a supermanifold $M$ is localized to a certain
subset of $M$.

In what follows $M$ is a compact
supermanifold  with $\dim M=(n_{+},  n_{-})$.
 For any vector field $F$ on $M$ we denote
by $R_{F} \subset M$ the zero locus of its number part $m(F)$.

\begin{thm}
Let $M$ be a compact supermanifold with a  volume
form $dV$.
Let $Q$ be an odd vector field on $M$ which satisfies the following
conditions:
\begin{eqnarray*}
(i)\  div _{dV} Q =0 \\
(ii)\  Q^2 \in {\cal K} (M)
\end{eqnarray*}
Then for any neighborhood
 $U( R_{Q})$  of $R_{Q}$ in $M$
there exists an even $Q$-invariant function $g_{0}$ which is equal to 1 on
$ \mbox{  some neighborhood } O(R_{Q}) \subset U(R_{Q})
\mbox{ of } R_{Q}$ and vanishes outside of
 $U(R_{Q})$ .   For
 every $Q$-invariant function $h \in C(M)$ and every $g_{0}$
obeying the above conditions we have
\begin{eqnarray}
\int _{M} dV h = \int _{M} dV g_{0} \cdot h
\end{eqnarray}
\end{thm}

The proof of Theorem 1 can be deduced from the following.

\begin{lm}
There exists an odd $Q^2$-invariant function $\sigma$ on $M$ which satisfies
$m(Q \sigma )(x) \neq 0 \mbox{ if } x \notin R_{Q}$.
\end{lm}

{\em Proof of the Lemma 1}.
Let us begin with preliminary remarks concerning the structure of the
supermanifolds.   Consider an ordinary manifold $N$ and a vector bundle
$\alpha$ over $N$.   Then we obtain a supermanifold from the total space
of $\alpha$ by reversing the parity of fibers.   This
manifold will be denoted by $\Pi \alpha N$ (a particular case of
this construction
was used in Sec.\ 2).
It is well known that every supermanifold can be obtained by
means of this construction \cite{Be}.

If we begin with a supermanifold $M$, then
 the construction of a bundle $\alpha$
over $N=m(M)$ with $M=\Pi \alpha (N)$ can
 be described in the following way:
let us fix an atlas of the supermanifold $M$.
Even local coordinates will be denoted by
Latin letters,  odd local coordinates will
 be denoted by Greek letters.
Let us represent the transition functions
from local coordinates
$(x^1,  \ldots,  x^{n_{+}} ;  \xi ^{1},  \ldots,  \xi ^{n_{-} })$ to local
coordinates
$(\tilde{x}^1,  \ldots,  \tilde{x}^{n_{+}} ;  \tilde{\xi } ^{1},
\ldots,  \tilde{\xi} ^{n_{-}} )$
as follows:
\begin{eqnarray}
\tilde{x} ^{i} =f^{i} (x) + \cdots \\
\tilde{\xi} ^{\alpha} =\phi ^{\alpha} _{\beta} (x)  \xi ^{\beta}+ \cdots,
\end{eqnarray}
where the omitted terms are of higher order with respect to the $\xi$'s.
The set of
functions $f^{i}$ can be considered as the set of transition functions
between local coordinate systems on the
body $N$ of $M$.   The functions
 $\phi ^{\alpha} _{\beta}$ are transition functions
of a bundle $\alpha$ over $N$.   One can prove that
$M$ is diffeomorphic to $\Pi \alpha (N)$.
There exists also an invariant construction of the bundle $\alpha$
(which is often refered to as the conormal bundle,  see \cite{Be}).

There is a natural homomorphism of the group
$Diff (M)$ of diffeomorphisms of $M$ into
the group of automorphisms of the bundle
$\alpha$ (the existence of such a homomorphism
follows immediately from the existence of  an
invariant construction of $\alpha$).   Even vector fields
on $M$ can be considered as infinitesimal diffeomorphisms of $M$.
Therefore they generate
infinitesimal automorphisms of the vector bundle $\alpha$.
In other words there exists a natural homomorphism
 of the Lie algebra of vector
fields on $M$ into the Lie algebra of infinitesimal automorphisms of the
vector bundle $\alpha$;  the infinitesimal automorphism corresponding
to the vector field $A$ will be denoted by $\overline{A}$.
In local coordinates $(z)=(x^i,  \xi ^{\alpha })$ on $M$
\begin{eqnarray}
Q=
\sum_{i=1}^{n_{+}} a_{\alpha}^{i} (z) \xi ^{\alpha}
\frac{\partial}{\partial x^i} +\sum_{\alpha =1}^{n_{-}}
b^{\alpha} (z) \frac{\partial}{\partial \xi ^{\alpha} },
\end{eqnarray}
where $a_{\alpha}^{i} (z)=a_{\alpha}^{i} (x) +\ldots$,
$b^{\alpha} (z)=b^{\alpha} (x)+\ldots$.  Here and below
we denote
the  higher order terms in $\xi$'s by $\ldots$.
Also,
$Q^2=\sum_{i=1}^{n_{+}} k^i (z)
\frac{\partial}{\partial x^i} +\sum_{\alpha =1}
^{n_{-}} l_{\beta}^{\alpha} (z) \xi ^{\beta}
\frac{\partial}{\partial \xi ^{\alpha} }$,
where $k^i (z)=k^i (x) + \ldots ,
l_{\beta}^{\alpha} (z)=l_{\beta}^{\alpha} (x)
+ \ldots$.   The coefficients
$k^{i} (z) ,  l^{\alpha} _{\beta} (z)$ can be easlily
expressed in terms of $a^{i} _{\alpha} (z),  b^{\alpha}
(z)$,  but we will not need explicit formulae.
It follows from
$[Q,  Q^2 ]=0$ that
\begin{eqnarray}
 k^i (x) \frac{\partial b^{\alpha} (x) }{\partial
 x^i } -l_{\beta}^{\alpha} (x) b^{\beta} (x) =0
\end{eqnarray}
It is easy to see that $b^{\alpha} (x)$ defines
a section of  the bundle $\alpha $;  by definition
this section coincides with the number part of $Q$.
The relation (9) shows
that this section is invariant with respect to the infinitesimal
automorphism $\overline{Q^2}$ of the bundle $\alpha$.
In the conditions of Theorem 1 we can assume without loss
of generality that the group generated by $Q^2$ is dense
in some compact group $G$ of transformations of $M$.   We
conclude from this fact and from $\overline{Q^2}$-invariance of
the section $b^{\alpha}$ that this section is also invariant
with respect to the natural action of the group $G$ on the bundle $\alpha$.

Consider now any odd function
$\sigma $ on $M$.   In local coordinates
$\sigma (z)= \sigma _{\alpha} (x) \xi ^{\alpha}
+\mbox{higher order
terms in $\xi 's$}$.   It is clear
therefore that $\sigma$ determines
a section of the vector bundle $ \alpha ^{*} $
dual to $\alpha $.
Let $g_{\alpha \beta} (x)$ be a $G$-invariant
non-degenerate scalar product on the fibers of $\alpha (m(M))$,
establishing an isomorphism $ \alpha (m(M)) \cong \alpha^{*}
(m(M))$.   Such a scalar product always exists, since
$G$ is compact.   Let us take an odd function
$\sigma $ on $M$  such that
\begin{eqnarray*}
\sigma (z )= g_{\alpha \beta } (x) b^{\alpha} (x)
\xi ^{\beta} +\ldots.
\end{eqnarray*}
One can assume that $\sigma$ is $G$-invariant.  (If $\sigma$ is
not $G$-invariant take its average with respect to $G$.   As
$b^{\alpha}$ and $g_{\alpha \beta}$ are $G$-invariant this
operation does not change the terms linear in $\xi$'s.)   Therefore
$Q^2  \sigma  =0$ and $m(Q \sigma)
(x) =g_{\alpha \beta} (x) b^{\alpha} (x)
b^{\beta} (x) \neq 0$,  whenever $x \notin R_{Q}$.
This completes the proof of the lemma.

The lemma admits a simple corollary.

Let us introduce the notation
\begin{eqnarray}
\beta =\frac{\sigma }{Q
\sigma },
\end{eqnarray}
where $\sigma$ is the function constructed above.
The odd function $\beta$ is defined
on the complement of $R_{Q}$ in $M$ and satisfies
there the condition $Q \beta =1$.
(To check the last fact note that $(Q \sigma ) \beta =\sigma$ and
apply $Q$ to both sides of this equation.)

Consider now an arbitrary neighborhood $U(R_{Q})$ of $R_{Q}$.
Using the function $\beta$ one can construct a partition of unity on
$M,  \sum_{n \in J}
g_{n}  =1,  $which satisfies the following
conditions:
\begin{eqnarray}
i)&&\ supp (g_{0} )
 \subset U(R_{Q}),  0 \in J,   \nonumber \\
ii)&&\ \mbox{There exists a neighborhood
 $O(R_{Q})$ of $R_{Q}$ such that}  \\
 &&O(R_{Q}) \subset U(R_{Q})
 \mbox{ and } g_{0} \mid  _{O(R_{Q})} =1 ,   \nonumber \\
iii)&&\ Q g _{n} =0,  n \in J,  g_{n}=Q \rho _{n},  n \neq 0 ,   \nonumber
\end{eqnarray}
where the $\rho_{n}$ 's are some odd functions on $M$.

{\bf Proof.   }One can choose a finite atlas
$\{ U_{n} ,  n \in I \}$ of $M$ such that $R_{Q} \subset
\bigcup_{n  \in I' \subset I} U_{n} \subset U(R_{Q})$ and
$O(R_{Q}) \bigcap (\bigcup_{n \in I \setminus I'} U_{n})
=\emptyset$.   Now take a partition of unity on $M$,
$\sum_{n \in I} f_{n}=1$,  such that $supp(f_{n})
\subset U_{n} ,  n \in I $.   Without loss of generality
we can consider this partition of unity to be
$G$-invariant (otherwise take its average with respect to the
action of group $G$).   Therefore
our partition of unity is also $Q^2$-invariant.
The partition of unity
having the desired properties is given by $\{g_{n},  n \in
 J \}$,  where$J=(I \setminus I')  \bigcup \{ 0\},
g_{n}  = Q (\beta
f_{n}) ,  n \in I \setminus I' ,  g_{0}=1-\sum_{n \in I
\setminus I'} g_{n}$.   To prove  this note that
by
construction $\sum_{n \in I
\setminus I'}
f_{n} (p) =0$ for $p \in O(R_{Q})$ and $\sum_{n \in I
\setminus I'}
f_{n} (p) =1$ for
$p \not \in \bigcup_{n \in I'} U_{n}$.
Then $\sum_{n \in  I \setminus I' } Q(\beta
f_{n})(p) =1$ for $p \not \in  \bigcup_{n \in I'}
U_{n}$ and $\sum_{n \in  I \setminus I' }
Q(\beta f_{n})(p)=0$ for $p \in R$.   Also,  $Q(Q(\beta
f_{n}))=0$,  as $Q^2
f_{n} =0,  n \in I$.

Thus the existence of the $Q$-invariant
 partition of unity  satisfying (11) is proved.

Now we are in position to
complete the proof of Theorem 1.   Namely,  consider an integral
$Z=\int_{M} dV h$,  where $h$ is any $Q$-invariant
function on $M$.   Using
the $Q$-invariant partition of unity satisfying (11) one can rewrite an
expression for $Z$ in the
following way:
\begin{eqnarray*}
Z=\sum_{n \in I \setminus I'} \int_{M} dV
Q(\rho _{n} h) +\int_{M} g_{0} h
\end{eqnarray*}

But $div_{dV} Q=0$,  therefore $\int_{M} dV Q(\rho _{n}h )
=0  \mbox{ for all } n \in I \setminus I'$.   So we conclude that
\begin{eqnarray}
Z=\int_{M} dV g_{0} h,  \\
supp(g_{0}) \subset U(R_{Q}) \subset M,
Q g_{0} =0,  g_{0} \mid _{O (R_{Q})} =1
\end{eqnarray}
The last thing to be proved is that the function $g_{0}$ entering
the partition of unity can be replaced by any function obeying
(13) without changing the value of the integral (12).
Suppose there is an even function
$\tilde{g} _{0}$ on M which obeys (x) with $O(R_{Q})$ replaced by
$\tilde{O} (R_{Q})~,  ~ R_{Q} \subset
 \tilde{O} (R_{Q} ) \subset U(R_{Q} )$.
But then $(g_{0} -\tilde{g} _{0} ) \mid
  _{O (R_{Q}) \bigcap \tilde{O} (R_{Q}) }
\equiv 0$,  therefore $g_{0}- \tilde{g}_{0}
=Q(\beta (g_{0} -\tilde{g}_{0}))$,
where $\beta$ is defined by the formula (x).
We arrive at
the desired result by means of the following simple calculation:
\begin{eqnarray*}
\int_{M}dV g_{0} h -\int_{M} dV \tilde{g} _{0} h =
\int_{M} dV Q(\beta (g_{0}-\tilde{g} _{0})) h=
\int_{M} dV Q(\beta h (g_{0} -\tilde{g}_{0} )) =0
\end{eqnarray*}

This completes the proof of Theorem 1.

One can weaken the conditions on $Q^2$ in the Theorem 1 in the following
way.   Notice that the
 condition $Q^2 \in {\cal K}(M)$
means that  there exists a compact group $G \subset Diff(M)$  such that
$Q^2$ can be represented in the form
\begin{eqnarray}
Q^2 = \sum_{i=1}^{dim  ~G} p_{i}
e_{i}~,  \qquad p_{i} \in {\bf R},
\end{eqnarray}
where $\{ e_{i} \} _{i=1}^{dim ~ G}$ is a basis
of  Lie algebra ${\cal G}$ of $G$.
It is always possible to choose $G=G_{Q^2}$(see the
Introduction).   Then the one parameter subgroup generated by $Q^2$
will be dense in $G$.

One can generalize Theorem 1,  assuming that the coefficients $p_{i}$
in (14) are arbitrary even functions on $M$.
In the proof of the Theorem 1 we
used the fact that  the $p_{i}$'s are constants only once - when we proved
that from $\overline{Q^2}$-invariance of the section
$b^{\alpha}$ we can derive the
invariance of this section with respect to the natural action
of the group $G$ on the bundle
$\alpha$.   However the $G$-invariance of the section
$b^{\alpha}$ can be proved
if we know just that the vector field $Q$ is $G$-invariant.

{}From this remark it becomes clear that one can prove the following statement:
\begin{thm}
Suppose that all conditions of Theorem 1 are satisfied
except $(ii)$.   Impose instead the following conditions
on $Q$ :
\begin{eqnarray*}
 a)&& Q^2=\sum_{i=1}^{dim ~G} p_{i}
e_{i};  \mbox{where $p_{i}$ are even functions on $M$.} \\
b)&& \mbox{The vector field $Q$ on $M$ is $G$-invariant. }
\end{eqnarray*}

Then the conclusion of Theorem 1 is true.
\end{thm}

It follows from Theorems 1 and  2 that the values of the function $h$ on the
complement to an arbitrary neighborhood of $R_{Q}$ are irrelevant in
the calculation of $\int_{M} dV h$.   One can express this statement by saying
that this integral is localized on $R_{Q}$.

Now let us discuss the localization of the integrals of functions invariant
with respect to several anticommuting odd symmetries.
We will prove the following.

\begin{thm}
Let $\{ Q_{i} \} _{i=1} ^{N} $ be  odd anticommuting volume preserving
vector fields on $M $(i.e.,  $ \{ Q_{i} ,  Q_{j} \} =0 \mbox{ for }
i \neq j,
div _{dV}  Q _{i} =0,      1 \leq  i,  j \leq N$).  Let us
assume that  $\{ Q_{i} ,  Q_{i} \}  \in {\cal K} (M)$
 and $R_{Q_{i}} =R_{Q_{i}^2},
1 \leq i \leq N$.

Then for every function $h$ on $M$,   obeying $Q_{1} h=Q_{2} h=
\ldots =Q_{N} h =0$,   the integral of $h$ over $M$ is localized
on $R_{Q_{1}} \bigcap R_{Q_{2}} \bigcap \ldots \bigcap R_{Q_{N}}$.
\end{thm}

The precise meaning of the word "localized" is explained above.

To prove Theorem 3 let us notice that it follows
from $\{ Q_{i} ,  Q_{j} \}=0$ that
$ [ Q_{i} ^2 ,  Q_{j} ^2 ] =0$.   Consider
the group $G_{i}=[e^{t Q_{i} ^2}]$ defined as the closure
of the one-parameter subgroup
$\{ e^{t Q_{i} ^2} \}  \subset
Diff(M)$ in the compact open topology on $Diff(M)$ (see Introduction).
Taking into account that $Q_{i} ^2$ commutes with $Q_{j} ^2$ we see that
the subgroups $G_{i}$ and $G_{j}$ of  $Diff(M)$ commute
(we use the fact that
the one-parameter subgroup  $\{ e^{(t Q_{i} ^2 } \} $
is dense in the corresponding group $G_{i}$).

This means that the group
$G =  G_{1} \times G_{2} \times \ldots \times G_{N}$
acts in a natural way on $M$:
\begin{eqnarray*}
&&G \times M \rightarrow M \\
&&((g_{1} ,  \ldots ,  g_{N} ),  x) \mapsto g_{1} \circ \ldots \circ g_{N}
(x)
\end{eqnarray*}
Consider the odd vector field $Q =\sum_{i=1}^{N} c_{i} Q_{i}$,
where $\{ c_{i} \} _{i=1}^{N}$ is a set of  real numbers.

It is easy to check  that $Q h=0,  div _{dV} Q=0$ and $Q^2=
\sum_{i=1}^{N} c_{i} ^2 Q_{i} ^2$.   Therefore,  $Q^2 \in
{\cal L}ie\ G$  and we conclude that  $Q^2 \in \cal{K} (M)$.

One can choose $\{ c_{i} \} _{i=1}^{N}$ in such a way that the
one-parameter
subgroup generated by    $Q^2$ is dense in $G$ .   Then
$R_{Q^2} =\bigcap _{i=1} ^{N}
R_{Q_{i}^2}=\bigcap _{i=1} ^{N} R_{Q_{i}}$.   Taking into account that
$R_{Q} \subset R_{Q^2}$,
we see that $R_{Q} \subset \bigcap _{i=1}^{N}
R_{Q_{i}}$.   But  $Q$ is a linear combination of $Q_{i}$'s,
therefore  $\bigcap _{i=1} ^{N} R_{Q_{i}} \subset R_{Q}$.
Thus  $R_{Q}$ coincides with $\bigcup _{i=1} ^{N}
R_{Q_{i}}$.   Using Theorem 1 we see that the integral $\int_{M} dV h$
is localized
to $R_{Q}=\bigcap_{i=1} ^{N} R_{Q_{i}}$.   This proves Theorem 3.

To conclude the present section let us make the following remark.
Theorems 1 and  2
state the localization of integrals of $Q$-invariant functions
on the zero locus $R_{Q}$ of the number
part of $Q$.
In the next section we will apply Theorem 1 to give conditions of
exactness of the stationary phase approximation.
We will see that under these conditions
one can make the stronger statement that the integrals
at hand are localized to the zero locus $K_{Q}$ of the vector field
$Q$.

 \section{Exactness of stationary phase approximation.}

 In the previous section we formulated sufficient conditions
for the localization of integrals over supermanifolds.
Here we will consider the problem of calculation of such
integrals.   Namely,
we are going to decribe a class of examples in which the  integral can be
{\em exactly} computed by means of the stationary phase method.

Throughout this section we consider the integral
\begin{eqnarray*}
Z=\int_{M} dV \cdot h,
\end{eqnarray*}
 where $h $ is a $Q$-invariant function on $M$ and
$M,  dV,  Q$ satisfy
conditions of Theorem 1.   This means that our integral is localized
on $R_{Q}$.   Denote the zero locus of $Q$ by $K_{Q}$.
We restrict ourselves to the situations when
$K_{Q}$ is either a finite subset of $M$ or a compact submanifold
of $M$.   Moreover, we'll assume that the odd codimension
of $K_{Q} \subset M$ is equal to its even codimension.   In
the case when $K_{Q}$ is a
finite subset of $M$ the last restriction means that
the even dimension of $M$ is equal to its odd dimension.

Let us begin with one important remark
which will be exploited throughout the rest of the paper.
Suppose one can find an odd function
$\sigma $ on $ M$ such that $ Q^2  \sigma =0$.
Then $Q(e^{i \lambda Q \sigma})=0$ and therefore
\begin{eqnarray*}
\frac{d}{d \lambda} \int_{M} dV h e^{i \lambda Q \sigma } =
\int_{M} dV Q(h e^{i \lambda Q \sigma}) =0,
\end{eqnarray*}
(To conclude that the last integral is zero we used the fact that
the volume element $dV$ is $Q$-invariant; i.e.,    $div_{dV} Q=0$).
We see that $\int_{M} dV h e^{i \lambda Q \sigma}$ does not
depend on $\lambda$.   Therefore,
\begin{eqnarray}
Z=\lim_{\lambda \rightarrow \infty } \int_{M} dV h e^{i \lambda Q
\sigma} .
\end{eqnarray}

In what follows we will use (15) with the function $\sigma$
which was constructed in the proof of Lemma.

First we will compute $Z$ for the case when
$K_{Q} \subset M $  is  finite.
As was mentioned above we assume in this situation that
the even dimension $n_{+}$ of $M$ is equal to its odd
dimension $n_{-}$.
It follows from the condition of finiteness that
actually $K_{Q} \subset m(M)$.
(To prove this, notice that if  the point $p\in M$ having local coordinates
$(x(p),  \xi (p))$ belongs to $K_{Q}$,
then any point with coordinates $(x(p),  c \xi (p))$,
where $c$ is a real number,   also does. )

Let $\{ z=(x,  \xi) \}$ be a local coordinate system
on $M$ centered at a fixed point $p$ such that $Q(p)=0$.   In
such coordinates
\begin{eqnarray}
&&Q=(b^{\alpha} (x) + b^{\alpha}_{\beta \gamma} (x)
\xi ^{\beta} \xi ^{\gamma} + \ldots ) \frac{\partial}
{\partial \xi ^{\alpha}} +(c^{i}_{\alpha} (x) \xi ^{\alpha}
+ \ldots )\frac{\partial}{\partial x^{i}} \\
&&b^{\alpha} (0) =0,       \nonumber
\end{eqnarray}
where
``$\ldots$'' denotes as always
higher order terms in $\xi$'s.
We  call this zero
non-degenerate if $det(\frac{\partial b^{\alpha}}
{\partial x^{i}} (0)) \neq 0,  det(c^{i}_{\alpha} (0))
\neq 0$ (recall that $n_{+} =n_{-}$).

Consider the odd function ${\sigma} $
constructed in the  proof of  the Lemma 1:
\begin{eqnarray}
\sigma  (x,  \xi )=
g_{\alpha  \beta} (x) b^{\alpha} (x) \xi ^{\beta}
+\ldots
\end{eqnarray}
Notice that $Q\sigma (x,  \xi)=g_{\alpha
 \beta }(x) b^{\alpha} (x) b^{\beta} (x)
 +d_{\alpha \beta} (x) \xi ^{\alpha }
 \xi^{\beta} +\ldots$,  where $d_{\alpha \beta}(x)$ is some matrix .
The condition $Q(Q\sigma) =0$ leads to
 the following relation between $d_{\alpha \beta } (0)$
and $b_{ij} (0) \equiv \frac{\partial}{\partial
 x^{i}} \frac{\partial}{\partial x^{j}}
(g_{\alpha \beta} b^{\alpha} b^{\beta}) (0)$:
\begin{eqnarray}
\frac{\partial b^{\alpha}}{\partial x^{j}} (0)d_{\alpha \beta} (0)
+c^{i}_{\beta} (0) b_{ij} (0) =0
\end{eqnarray}
{}From (18),  the  nondegeneracy of the zero of the vector
field $Q$ at $p$ and the nondegeneracy of scalar
product $g$
it follows that the matrix $d_{\alpha \beta } (0)$ is also non-degenerate.
Therefore the point $p \in K_{Q}$ is a
non-degenerate isolated critical point
of the the function $Q \sigma$.   Consequently,
there exists a neighborhood $U(p)$ of the point $p$  such that $p$
is the only critical point of $ Q\sigma$ restricted to $U(p)$.
It also follows from (18) that

\begin{eqnarray}
sdet(Hess Q \sigma (p))
=sdet (-I Q'(p)),
\end{eqnarray}
where $I=antidiag(1,  1,  \ldots,  1) $ is $2n \times 2n$ matrix ,
$Q'(p)$ is the (super)matrix of first derivatives
of coefficient functions of operator $Q$ at the
point $p$.   Let us note that the condition
of non-degeneracy given above can be reformulated
in terms of $Q'$: a point $p$ is a non-degenerate zero
of $Q$ if $I Q'(0)$ is non-degenerate as a supermatrix
{}.
Finally,  the answer for $Z$ can be obtained by means
of the following calculation:
\begin{eqnarray}
Z \equiv \int_{M} dV h \stackrel{\rm Thm. 1}{=}
\int_{M} dV g_{0} h \stackrel{(15)}{=}
\lim_{\lambda \rightarrow \infty} \int_{M} dV
g_{0} h e^{ i \lambda Q \sigma} = \nonumber \\
\sum_{p \in K_{Q}} \frac{\rho (p) h(p)}
{\sqrt{sdet (Hess(Q \sigma)(p))}}
\stackrel{(19)}{=}
 \sum_{ p \in K_{Q}} \frac{\rho (p) h(p)}
{\sqrt{sdet (-IQ'(p))}},
\end{eqnarray}
where $supp(g_{0}) \subset \bigcup _{p \in K_{Q}}
U(p),  g_{0} \mid
 _{R_{Q}} =1$
 and $U(p)$ are chosen in such a way that the intersection
of the critical set of $\sigma$ with
$\bigcup _{p \in K_{Q}} U(p)$ is just $K_{Q}$;
$\rho (p)$ is the volume density at $p$.   The third equality in (20)
is due to the formula generalizing the
stationary phase approximation to the supercase
(see e.   g.   \cite{S1} for details).

Let us make a few remarks about (20).   First,  as it should be,
the final answer for $Z$ doesn't depend on the choice
of the non-degenerate inner product $g_{\alpha \beta}$ in
$\alpha (m(M))$.
Second,  notice that (20) states that $Z$ is represented as
a sum of contributions
of the zeros of $Q$;  this
is a stronger statement
than the one given by the general theorems of the previous section.

 Finally, consider the integrand of $Z$ in the form
 $h= e^{i S}$,  where $S $ is an even $Q$-invariant function.   It
is easy to check that $K_{Q}$ is contained in the critical
set of $S$. (This fact will be proved below in a more general
situation.)
Suppose also that $Hess (S)$ is non-degenerate at every $p \in K_{Q}$.
Then
one can rewrite (20) as follows:
\begin{eqnarray}
Z =\sum_{p \in K_{Q} } \frac{\rho (p) e^{i S(p)}}
{\sqrt{sdet (Hess(S)(p))}}.
\end{eqnarray}
The stationary phase approximation leads to  precisely such an expression
for $Z$, but with $p$ running through the critical set $K_{S}$ of $S$.
Formula (21)
means that the stationary approximation is exact and the points of
$K_{S} \setminus K_{Q}$ do not contribute to $Z$.

Now we will discuss more
general conditions on $h$ and $Q$ leading to the exactness of
the stationary phase
approximation for $Z$.

Our considerations are based on the following  statement:
\begin{lm}
Let
$h$ be a $Q-$invariant function on $M$ such that
$h \mid _{K_{Q}} =0$.   Let us assume that $K_{Q}$ is
a compact submanifold of $M$ and suppose $Q$ is non-degenerate
on $K_{Q}$; i.e.,
 the supermatrix $\partial _{\perp} Q (p)$ of transversal
derivatives of $Q$ at every $p \in K_{Q}$ is
invertible.
Then
\begin{eqnarray}
\int_{M} dV h =0.
\end{eqnarray}
\end{lm}

Let us introduce the following system of local coordinates
in a neighborhood of $K_{Q}$:
let  $\{x^{i};  \xi^
{\alpha} \}=\{ x^{i'},  x^{i ''};  \xi ^{\alpha '},
\xi ^{\alpha ''} \},  1 \leq i \leq n_{+} ,  1 \leq \alpha \leq n_{-},
1 \leq i' \leq n'_{+},  \leq \alpha ' \leq n'_{-},
1 \leq i'' \leq n''_{+},
1 \leq \alpha '' \leq n''_{-},  n_{+}=n'_{+} +n''_{+},
n_{-}=n'_{-} +n''_{-}$,  be such that $K_{Q}$ is singled out by the equations
$x'=0$,  $\xi ' =0$.
In other words the indices labeled by $'$ are related to
transversal directions  and tangent indices are labeled by $''$.
Note that by our assumption $n'_{+} =n'_{-}$.
Let us also introduce cumulative notations for
even and odd local coordinates in the vicinity of
$K_{Q}:\{z^{I'} \}=\{x^{i'},  \xi^{\alpha '} \},
\{z^{I''} \}=\{x^{i''},  \xi^{\alpha ''} \}$.
In this notation $K_{Q}$ is singled out by the equation $z'=0$.

We begin the proof with the following remark.
Consider any $Q$-invariant function $S$ on $M$ which is
locally constant on $K_{Q}$ (i.e., it is  constant on each connected
component of $K_{Q}$).   Then
$K_{Q} \subset
K_{S}$,  where $K_{S}$ is the critical set of $S$.
To see this it is sufficient to present $S$ in the vicinity of
$K_{Q}$ as a power series in $(z')$'s and impose the
condition $QS=0$.   As a consequence of the nondegeneracy of
$Q$ on $K_{Q}$
we will obtain $\frac{\partial S}{\partial z^{I'}} \mid _{K_{Q}} =0$.
Also $\frac{\partial S}{\partial z^{I''}} \mid _{K_{Q}}=0$ as $S$ is locally
constant on $K_{Q}$.
Therefore the inclusion $K_{Q} \subset K_{S}$ is proved.

Consider now the odd function $\sigma $ constructed in Lemma 1.
By construction $Q \sigma \mid _{K_{Q}}=0$,  $Q(Q\sigma)
=0$.   Therefore in the vicinity of $K_{Q}$
we have
\begin{eqnarray}
 Q\sigma (z'',  z') = S_{I',  J'} (z'') z^{I'} z^{J'}
+\mbox{higher order terms in $(z')$'s}.
\end{eqnarray}
To deduce the statement of Lemma 2 we notice that
\begin{eqnarray}
\int_{M} dV \cdot h \stackrel{\rm Thm. 1}{=} \int_{M}
dV \cdot g_{0} h \stackrel{(15)}{=} \lim_{  \lambda \rightarrow \infty}
\int_{M} dV \cdot
g_{0} h e^{i \lambda Q \sigma}.
\end{eqnarray}
The last expression in (24) is equal to zero;
we obtain this result from the standard formula
for the stationary phase approximation.   It is important to stress
that such a formula can be applied because (as we are going to prove
below) the Hessian of $\lambda Q \sigma$ in the
directions transverse to $K_{Q}$ is non-degenerate.
We don't need an exact formula for stationary phase approximation;
we use only that the leading term in this approximation is of order
$\lambda ^{0}$(because $n_{+} ' =n_{-} '$) and that the answer
is proportional to
$h \mid _{K_{Q}}=0$.

So,  let us prove the nondegeneracy of the
Hessian of  $\lambda Q \sigma$
in the directions transversal to $K_{Q}$. At the point of $K_{Q}$ with
coordinates $z''$ this Hessian coincides with the supermatrix
$S_{I',  J'} (z'')$  entering (23).
First we will prove that the even-even and odd-odd parts of
$S_{I',  J'}$ are
degenerate or non-degenerate simultaneously.
Then we will show that the even-even
part of  $S_{I',  J'}$ is non-degenerate.

In the vicinity of $K_{Q}$ we have
\begin{equation}Q= a^{I'}_{J'} (z'',  z') z^{J'}
\frac{\partial}{\partial z^{I'}} +
b^{I''}_{J'} (z'',  z') z^{J'} \frac{\partial}
{\partial z^{I''}}\nonumber
\end{equation}
The nondegeneracy of $Q$ in a
neighborhood of $K_{Q}$ means that the matrices
$a^{i'}_{\lambda'} (z'',  0)$,  $a_{i'}^{\lambda'}
(z'',  0)$ are invertible.
The condition $Q(Q\sigma) =0$ leads in particular
to the following
condition on $S_{I',  J'}$:
\begin{eqnarray}
\lefteqn{a^{i'}_{j'} (z'',  0) S_{i',   \gamma} (z'')+
a^{\delta '}_{\gamma '} (z'',  0) S_{j',   \delta '} (z'')+}\nonumber\\
&&a^{\delta '}_{j'} (z'',  0) S_{\delta ',   \gamma '} (z'')+
a^{i'}_{\gamma '} (z'',  0) S_{i',  j'} (z'')=0
\end{eqnarray}
We will need a corollary of (26) for the number parts of matrices
which enter it.
The odd matricies  $a^{i'}_{j'} (z'',  0)$ and
$a^{\delta '}_{\gamma '} (z'',  0)$  have zero number parts,
therefore it follows from (26) that
\begin{eqnarray}
m(a^{\delta '}_{j'} (z'',  0)) m( S_{\delta ' ,  \gamma '} (z''))+
m(a^{i'}_{\gamma '} (z'',  0)) m( S_{i',  j'} (z''))=0
\end{eqnarray}
{}From the invertibility of
$a^{i'}_{\lambda '} (z'',  0),  a_{i'}^{\lambda '}
(z,  0)$  it follows that $\det (m(a^{i'}_{\lambda '} (z'',  0)))
\neq 0$,   $\det (m(a_{i'}^{\lambda '} (z'',  0))) \neq 0$.   Therefore, in
view of (27), the matrices
$m( S_{\delta ' ,  \gamma '} (z''))$,
$m( S_{i',  j'} (z''))$ are singular or non-singular simultaneously.
We will prove that
$\det (m( S_{i',  j'} (z'')) ) \neq 0$.   Then by (27)
$\det (m( S_{\delta ',   \gamma '} (z''))) \neq 0$.
Invertibility of the number parts of matrices depending on odd variables
leads to the invertibility of matrices themselves;  therefore
we conclude from the above considerations that the matrices
$S_{i' j'} (z'')$,  $S_{\delta ' \gamma '} (z'')$ are
invertible.

So it remains to prove invertibility of the number part
$ m( S_{i',  j'} (z''))$
of the even-even block $S_{i'j'}(z'')$
 for all $z''$'s.   By construction,   $m(Q \sigma (x,  \xi))=
g_{\alpha \beta} (x) b^{\alpha}(x) b^{\beta}(x) $,
where $g$ is a $\overline{Q^2}$-invariant
scalar product in $\alpha (m(M))$.
Therefore
\begin{eqnarray}
 m( S_{i',  j'} (z''))=
g_{\alpha \beta} (x'') b^{\alpha}_{i'}(x'') b^{\beta}_{j'} (z''),
b_{i'}^{\alpha} (x'') =\frac{\partial}{\partial x^{i'}}
b^{\alpha} (x)\mid _{x'=0}
\end{eqnarray}
It follows from  $L_{\overline{Q^2}} g=0$
that
\begin{eqnarray}
( L_{\overline{Q^2}} g)_{\alpha \beta} (x)=(
k^{i}  \frac{\partial}{\partial x^{i}} g_{\alpha \beta}
+g_{\gamma \beta } l^{\gamma}_{\alpha}+g_{\gamma \alpha}
l^{\gamma}_{\beta}) (x) =0,
\end{eqnarray}
where $ k^{i},  l_{\alpha} ^{\beta} $ are
the coefficients of $Q^2$ introduced in the text following (8).
Consider (29) in such coordinates in the vicinity of
the point $(z'',  0)$ that $g_{\alpha \beta} (z'',  0) =
\delta _{\alpha \beta}$.   Then (29) written at the point
$(z'',  0) \in K_{Q}$ gives
\begin{eqnarray*}
a_{\delta}^{i'} \frac{\partial}{\partial x^{i'}}
b^{\gamma}+
a_{\gamma}^{i'} \frac{\partial}{\partial x^{i'}}
b^{\delta} =0
\end{eqnarray*}

Noticing that $Q \mid _{K_{Q}} =0$,  we get  $a_{\delta ''}^{i'}
(z'',  0) =0 $ for all tangent indices $\delta ''$.   Then it
follows from the last equation that
$a_{\gamma '}^{i'} b^{\delta ''}_{i'} \mid_{K_{Q}}
=0 $.   But
$\det(a_{\gamma '}^{i'})\mid _{K_{Q}} (x)
\neq 0$ by nondegeneracy of $Q$ in vicinity
of $K_{Q}$.   Therefore in the given coordinates
 $b^{\delta ''}_{i'} \mid_{K_{Q}} =0$.   Therefore
by (28) $m((Q \sigma )_{i',  j'}) \mid _{K_{Q}}=
(b^{k'}_{i'} \cdot b^{k'}
_{j'}) \mid_{K_{Q}}$,  which is non-singular
(recall that nondegeneracy of $Q$ means that
$b^{k'}_{i'}$
is non-singular ).
Therefore Lemma 2 is proved.

Lemma 2 admits a simple corollary which is important for our
further considerations:
\begin{lm}
Let $S$ be an even $Q$-invariant function  where
$Q$ satisfies all conditions of Lemma 2.
Suppose    $S \mid _{K_{Q}} =0$.
Then $\int_{M} dV e^{i \beta S}$ does not depend
on the parameter $\beta$ and can be calculated therefore
according to the following formula
\begin{eqnarray}
\int_{M} dV =\lim_{\beta \rightarrow \infty } \int_{M} dV e^{i \beta S}
\end{eqnarray}
\end{lm}
Really,  since $S \mid _{K_{Q}}=0$,
$\frac{d}{d \beta} \int_{M} dV  e^{i \beta S} =
\int_{M} dV S e^{i \beta S} =0$ in accordance with Lemma 2 applied
to the integral $Z$ with the integrand $h=S e^{i \beta S}$.

Before proceeding further let us explain what we mean by saying that the
stationary phase approximation for the integral
$\int_{M} dV e^{i \beta S}$ is
exact.   Consider an asymptotical expansion in powers
of $\beta ^{-1}$ for such an integral.
Let us present the critical set of $S$ as a union of
level sets of $S$ on $K_{S}$;
$K_{S} =K_{S} ^{(1)} \bigcup K_{S} ^{(2)} \bigcup \ldots \bigcup K_{S} ^{N}$;
i.e.,
$S_{i} \equiv S \mid _{K_{S} ^{i}}$
is constant and $S_{i} \neq S_{j}$ if
$i \neq j$(each set $K_{S} ^{(i)}$ is a union
of connected components).   Then the asymptotical expansion of
the integral at hand takes the form
\begin{eqnarray}
\int_{M} dV e^{i \beta S} =\sum _{i=1}^{N} e^{i \beta S _{i}}
( c^{i} _{- \Delta} \beta ^{\Delta} +
c_{-\Delta +1} \beta ^{-\Delta +1} + \ldots +
c_{0} ^{i} +c_{1} ^{i} \beta ^{-1}  +\ldots),
\end{eqnarray}
where $\Delta$ is some number equal to the difference
between the odd and even codimensions
of $K_{S}$ in the case when the latter is a submanifold of $M$;
$c$'s are some constants.
We say that the stationary phase approximation for the integral is exact
if   the equation
$\int_{M} dV e^{i \beta S} =\sum_{j=1}^{N} e^{i \beta S_{j}} c_{0} ^{j}$
is satisfied (in particular this means that
$c_{k}^{i}=0$ for  $1 \leq i \leq N$ and $k \neq 0$).

Based on Lemma 3 one can easily prove the following Theorem:
\begin{thm}
Let $S$ be an even $Q$-invariant function on $M$ which is locally
constant on $K_{Q}$.   Suppose $Q$ is non-degenerate in a neighborhood
of $K_{Q}$.   Then the stationary phase approximation of the integral
$\int_{M} dV e^{i \beta S}$ is exact.
\end{thm}

We proved already that under the conditions
of Theorem 4, $K_{Q} \subset K_{S}$.   Let us decompose $K_{Q}$
into a union of level sets
$K_{Q} =\bigcup_{j =1} ^{L} K_{Q} ^{j},  L \leq N$.
In other words $K_{Q}^{j} =K_{S}^{j} \bigcap K_{Q}$.

Using Theorem 1 let us present the integral under consideration in the form
\begin{eqnarray}
\int_{M} dV e^{i \beta S} =\sum_{j=1}^{L} e^{i \beta S _{j}}
\int_{M} dV g_{0} ^{(j)} e^{i \beta (S-S _{j})},
\end{eqnarray}
where $S_{j} \equiv S \mid _{K_{Q} ^{j}}$ and $g_{0} ^{(j )}$
is an even $Q$-invariant function from Theorem 1 having
support in some neighborhood of $K_{Q} ^{j}$.
By virtue of Lemma 3, the integrals in the right hand side
of (32) do not depend on $\beta$,

since $S-S_{j}$ vanishes on  $ K_{Q} ^{j}$.

Therefore the stationary phase approximation for
the integral $\int_{M} dV e^{i \beta S}$ is
exact.   This proves Theorem 4.

We proved even a stronger result:
the coefficients
$c_{k}^{i} $ in the decomposition (31) vanish if  $K_{S}^{j} \bigcap K_{Q}$
is empty.   This means that the asymptotical expansion for
$\int_{M} dV e^{\beta S}$
receives a contribution only from  part of the
critical  set $K_{S}$, namely
from the set $K_{Q}$.

Let us also notice that all statements in the present section
can be generalized to the case when
$K_{Q}$ is not a submanifold of $M$,
but is a union of  compact submanifolds ,  not
necessarily of the same dimension.

\section{Appendix.}

In this paper we utilize more or less the standard
terminology of supergeometry (see for example
\cite{Be}, \cite{S2}).
We will use the definition of  an $(m \mid n )$-dimensional
supermanifold as an object obtained from domains
in  $(m \mid n)$-dimensional superspace $R^{(m \mid n)}$
pasted together by means of invertible maps.
(See for example \cite{S2} for a more precise definition.)
The body $m(M)$ of the supermanifold $M$ can be identified with
the submanifold of $M$ singled out by the equations
$\xi ^{1} =\xi^{2}=\ldots = \xi^{n} =0$,
where $\xi^{1},\xi^{2}, \ldots ,\xi^{n}$ are odd coordinates.
This condition is independent of the choice of coordinate system
because we do not consider families of supermanifolds and therefore
the transition functions between different local coordinate
systems do not depend on external odd parameters.

With every $(m \mid n)$-dimensional supermanifold $M$ one can associate
an $n$-dimensional vector bundle over $m(M)$,
the so called conormal bundle (see \cite{Be} for
an invariant definition and Section 3 for the coordinate construction).
For every function $F$ on a supermanifold $M$ we can consider
its number part $m(F)$ as a restriction of  $F$ to the body
$m(M)\subset M$.    If $A$ is an even vector field then
one can define its number part as a vector field on $m(M)$.
If in local coordinates the vector field $A$ corresponds to
a first order differential operator
\begin{eqnarray*}
A=A^{i} (x,\xi) \frac{\partial}{\partial x^{i}} +
A^{\alpha} (x, \xi) \frac{\partial}{\partial \xi ^{\alpha}},
\end{eqnarray*}
then its number part corresponds to an operator
\begin{eqnarray*}
m(A)=A^{i} (x,0) \frac{\partial}{\partial x ^{i}}.
\end{eqnarray*}
The number part of an odd vector field $Q$ is defined as a
section of the conormal bundle.   If in local coordinates
\begin{eqnarray*}
Q= {\kappa}^{i} (x, \xi) \frac{\partial}{\partial x^{i}} +
q^{\alpha} (x,\xi) \frac{\partial}{\partial \xi ^{\alpha}} ,
\end{eqnarray*}
then this section is specified by means of functions
$q^{1} (x,0),q^{2} (x,0),\ldots q^{n} (x,0)$.

Let us finally mention that we denote by $sdet~ M$ the superdeterminant
(Berezinian) of a supermatrix $M$.The supermatrix  $M$ is invertible
iff  the even-even and odd-odd blocks of $M$ are non-singular in
the usual sense;
then $sdet~M$ exists and its number part does not vanish.

{\em Acknowledgements.} We are grateful to D. Fuchs, A. Polyakov and
E. Witten for useful discussions.Special thanks are due to M. Penkava for
reading and editing the manuscript.

\end{document}